\documentclass[aps,prd,amssymb,eqsecnum]{revtex4}
\usepackage{graphicx}% Include figure files
\usepackage{bm}% bold math

\begin{document}

\title{Two approaches for the gravitational self force in black hole
spacetime:\\ Comparison of numerical results}
%\title{Gravitational self-force in circular orbits around a Schwarzschild
%black hole: Agreement between computations in Lorenz and Regge-Wheeler gauges}

\author{Norichika Sago$^{1}$, Leor Barack$^{1}$ and Steven Detweiler$^{2}$}
\affiliation
{$^1$School of Mathematics, University of Southampton, Southampton,
SO17 1BJ, United Kingdom\\
$^2$Institute for Fundamental Theory, Department of Physics, University of Florida,
Gainesville, FL 32611-8440}

\date{\today}

\begin{abstract}

Recently, two independent calculations have been presented of finite-mass
(``self-force'') effects on the orbit of a point mass around
a Schwarzschild black hole. While both computations are based on the
standard mode-sum method, they differ in several technical aspects, which makes
comparison between their results difficult---but also interesting.
Barack and Sago [Phys.\ Rev.\  D {\bf 75}, 064021 (2007)] invoke the notion of a
self-accelerated motion in a background spacetime, and perform a direct calculation
of the local self-force in the Lorenz gauge (using numerical evolution of
the perturbation equations in the time domain); Detweiler [Phys.\ Rev.\ D {\bf 77},
124026 (2008)] describes the motion in terms a geodesic orbit of a (smooth)
perturbed spacetime, and calculates the metric perturbation in the Regge--Wheeler
gauge (using frequency-domain numerical analysis).
Here we establish a formal correspondence between the two analyses, and
demonstrate the consistency of their numerical results.
Specifically, we compare the value of the conservative
$O(\mu)$ shift in $u^t$ (where $\mu$ is the particle's mass and $u^t$ is the
Schwarzschild $t$ component of the particle's four-velocity), suitably mapped
between the two orbital descriptions and adjusted for gauge.
We find that the two analyses yield the same value for this shift within mere
fractional differences of $\sim 10^{-5}$--$10^{-7}$ (depending on the orbital
radius)---comparable with the estimated numerical error.

\end{abstract}

%\pacs{04.25.Nx, 04.30.Tv, ???}
% 04.25.Nx  Post-Newtonian approximation; perturbation theory;
%           related approximations
% 04.30.Db  Wave generation and sources
% 04.30.Nk  Wave propagation and interactions
% 04.30.Tv  Gravitational-wave astrophysics
%           (see also 95.85.Sz Gravitational radiation,
%            magnetic fields, and other observations in astronomy)
% 04.80.-y  Experimental studies of gravity
% 04.80.Nn  Gravitational wave detectors and experiments
%
% 07.05.-t  Computers in experimental physics
% 07.05.Kf  Data analysis: algorithms and implementation;
%           data management
%
%
% 95.30.Sf  Relativity and gravitation
%           (see also section 04 General relativity and
%            gravitation; 98.80.Jk Mathematical and relativistic
%            aspects of cosmology)
% 95.55.-n  Astronomical and space-research instrumentation
% 95.55.Ym  Gravitational radiation detectors; mass spectrometers;
%           and other instrumentation and techniques
%
% 95.85.-e  Astronomical observations
%           (additional primary heading(s) must be chosen
%            with these entries to represent the astronomical
%            objects and/or properties studied)
% 95.85.Sz  Gravitational radiation, magnetic fields,
%           and other observations
%
%\keywords{gravitational waves: inspiraling compact objects
% --- gravitational waves: theory --- data analysis}

\maketitle

\section{Introduction} \label{Sec:I}

Two recent works discussed the effects of gravitational self-force (GSF)
on the motion of a point particle in a circular orbit around a Schwarzschild
black hole. Barack and Sago \cite{Barack:2007tm} (hereafter BS) carried out
a direct calculation of both dissipative and conservative components of the
local GSF in the Lorenz gauge. Detweiler \cite{Detweiler:2008} (hereafter SD)
focused on quantifying a number of gauge-invariant orbital effects associated with
the finite mass of the particle. Both analyses employ the mode-sum method
\cite{Barack:1999wf,Barack:2001gx}, wherein the
metric perturbation associated with the particle is first decomposed into
multipole harmonics, and the contribution to the physical back-reaction force
is then calculated mode by mode. However, the two analyses use very difference
languages in describing the motion of the particle. BS's description adheres to
the original fundamental formulation of Mino, Sasaki and Tanaka \cite{Mino:1996nk}
(also Quinn and Wald \cite{Quinn:1996am} and the recent Gralla and Wald \cite{Gralla}),
wherein the GSF is viewed as accelerating the particle with respect to a background
geodesic. SD, instead, depicts the particle's orbit as a geodesic of a smooth,
perturbed spacetime---an interpretation first put forward by Detweiler and Whiting
\cite{Detweiler:2002mi}. The two descriptions of the motion are known to be
fundamentally equivalent \cite{Detweiler:2002mi} (see \cite{Poisson:2003nc} for
a pedagogical review), but the different language they use makes it rather
difficult to compare between the results of the respective calculations.

The goal of this report is to establish a ``common language'' to facilitate
comparison between the results of BS and SD, hence show that the two sets of
results are in full agreement with each other. The motivation for such a
comparison is three-fold: Firstly, it provides a reassuring confirmation for
the equivalence between the two descriptions of the motion. Secondly, it
demonstrates how the correspondence between the two descriptions
is to be established in practice---a similar technique could then
potentially be implemented in future calculations of the GSF. Thirdly,
 the comparison provides a good overall check
of both calculations. It is indeed a good check, because the two analyses are
highly independent: Not only do they invoke different interpretations of the
perturbed motion, they also use different gauges for the metric perturbation,
and apply very different numerical methods to evaluate it.

The GSF itself is gauge dependent \cite{Barack:2001ph}, and, of course, so is the
metric perturbation. To compare the results of two calculations held in different
gauges, one naturally seeks to consider gauge-invariant quantities which are
affected by the finite mass of the particle in a non-trivial way. A list of such
quantities, for circular
orbits in Schwarzschild, was introduced by Detweiler in \cite{Detweiler:2005kq}
(cf.\ SD). In discussing gauge-invariant GSF effects,
it is useful to distinguish between effects which are {\it dissipative} and
others which are {\it conservative}. Dissipative effects are manifest in a gradual,
secular drift in the value of the intrinsic orbital parameters (such as the energy
and angular momentum) due to the emission of gravitational waves. In the case of
a circular equatorial orbit in Schwarzschild, such effects arise from the $t$
and $\varphi$ components of the GSF (henceforth $t,r,\theta,\varphi$ denote the
standard Schwarzschild coordinates), which are directly related to the rate of
change of orbital energy and angular momentum, respectively. The deficit in
energy and angular momentum is carried away by the gravitational waves, and
can in principle be measured by a distant observer---it is therefore not surprising
to find that, in the above orbital setup, the $t$ and $\varphi$ components of the GSF
are {\it gauge invariant} (within a class of physically reasonable gauges).
This makes it relatively easy to test the values of these components in
numerical calculations. Indeed, both BS and SD were able to demonstrate that
their numerical results for the $t$ and $\varphi$ components of the local GSF
were in agreement with the values inferred from the fluxes of energy
and angular momentum in the gravitational waves (the latter values were derived
by BS and SD from their respective numerically-calculated metric perturbations,
and confirmed against values tabulated in the literature). Hence, BS and SD's
results for the dissipative part of the GSF are already well tested,
and we shall not consider them further in this work.

The situation with the conservative part of the GSF is more involved.
Conservative GSF effects are manifest in a small shift in the instantaneous
values of the
orbital parameters away from their unperturbed values; such effects alter
the time dependence of the orbital phases, and may have important influence
on the phasing of the emitted gravitational waves \cite{Pound&Poisson1,Pound&Poisson2}.
In the case of a circular orbit in Schwarzschild which concerns us here, the
conservative effects are entirely due to the radial ($r$) component of the GSF.
This component is not gauge invariant, and its value depends on the gauge in
which the associated metric perturbation is expressed. A meaningful,
gauge-invariant description of the conservative effects requires knowledge of
both the GSF and the metric perturbation associated with it. It is not possible
to check the self-consistency of the numerical calculation of the conservative
GSF's component using energy-momentum balance considerations as done with the
dissipative components. It is possible to test this calculation against results
from the post-Newtonian (PN) literature (as done in SD by examining gauge-invariant
quantities; see below), but such tests are only applicable for orbits with
a sufficiently large radius. The two independent analyses by BS and SD now provide
us with an opportunity to test the calculation of the conservative GSF effects
in the strong-field regime.

For our comparison we shall consider the two physically observable gauge-invariant
quantities identified by Detweiler in \cite{Detweiler:2005kq,Detweiler:2008}.
(The meaning of ``gauge invariance'' here will be made precise below; it refers to
a class of gauge transformations compatible with the helical symmetry of the
perturbed spacetime.) The first quantity is the azimuthal orbital frequency $\Omega$
(defined with respect to Schwarzschild time $t$), from which one derives a
``gauge-invariant'' orbital radius $R$ [see Eq.\ (\ref{R}) below]. The second gauge
invariant is $U\equiv u^t$, the contravariant $t$ component of the particle's
four-velocity.
%The values of both
%$u^t$ and $\Omega$ (or $R$) in the perturbed spacetime depend on the conservative
%GSF in a non-trivial way.
The functional form of the relation $U(R)$ is
independent of the gauge, and thus provides a convenient handle
with which to compare between two calculations held in different gauges.

The relation $U(R)$ (including finite-mass terms) has been calculated in SD and used there
as a basis for comparison with results from PN calculations. Here we will calculate
$U(R)$ based on the Lorenz-gauge results of BS. We shall see that a meaningful
comparison between $[U(R)]_{\rm SD}$ and $[U(R)]_{\rm BS}$ requires two important
adjustments: Firstly, one needs to account for the fact that the orbits in
the two analyses are formally defined in two different geometries: It is an
accelerated trajectory on a background spacetime in BS, whereas in SD it is
a geodesic in a perturbed spacetime. We shall see that a formal connection between
the two descriptions can be established by suitably relating the proper-time parameters
along the two orbits.

The second necessary adjustment is more subtle: The (perturbed) quantity $U$ was
shown in SD to be gauge invariant, for circular orbits in Schwarzschild, within
a class of ``physically reasonable'' gauges (we shall define this class more
accurately in the next Section). We will show, however, that the Regge-Wheeler-gauge
perturbation of SD and the Lorenz-gauge perturbation of BS are related by a
gauge transformation which does {\em not} leave $U$ invariant. We will then
work out explicitly a gauge transformation which brings the BS perturbation
into the same class of ``physically reasonable'' gauges as the SD perturbation,
and use this to gauge-adjust the value of $U$ (the necessary gauge adjustment
will have a monopole component only).
%As an aside, we will discuss the extension
%of the notion of ``physically reasonable'' gauges to a wider class which includes
%both BS's and SD's metric perturbations.

The structure of this work is as follows.
In Sec.\ \ref{Sec:II} we review BS's and SD's formulations of the orbital motion
and introduce the necessary notation for our analysis.
In Sec.\ \ref{Sec:III} we establish a formal relation between $[U(R)]_{\rm SD}$
and $[U(R)]_{\rm BS}$ by performing the two adjustments mentioned above.
In Sec.\ \ref{Sec:IV} we describe the numerical derivation of $U(R)$ within the
BS analysis, and compare the numerical values (adjusted for proper time and for gauge)
with those derived in SD. This comparison is displayed in Table I.

Throughout this work we use standard geometrized units (with $c=G=1$) and metric
signature $({-}{+}{+}{+})$.  $t,r,\theta,\varphi$ represent the standard
Schwarzschild coordinates.

%**************************************************************
\section{Review and Notation}\label{Sec:II}
%**************************************************************

BS and SD use rather different notation. For our purpose it will be
useful to introduce a unified notation, which is one of the aims of the
following short review.
Our convention will be that quantities arising from the BS analysis are
denoted with a `tilded' symbol (e.g., $\tilde X$), while their SD counterparts
are left `un-tilded':
%~~~~~~~~~~~~~~~~~~~~~~~~~~~~~~~~~~~~~~~~~~~~~~~~~~~~~~~~~~~~~~~~
\begin{eqnarray} \label{BSSD}
\tilde X: &\text{quantities arising from BS}, \nonumber\\
       X: &\text{quantities arising from SD}.
\end{eqnarray}
%~~~~~~~~~~~~~~~~~~~~~~~~~~~~~~~~~~~~~~~~~~~~~~~~~~~~~~~~~~~~~~~~
We will generally denote ``background'' quantities (i.e., ones obtained
when the metric perturbation and all GSF effects are ignored) with a
script `$0$' (as in $X_0$). To ensure that BS and SD
correspond to identical physical setups, we will require that all background
quantities attain the same values in both analyses:
%~~~~~~~~~~~~~~~~~~~~~~~~~~~~~~~~~~~~~~~~~~~~~~~~~~~~~~~~~~~~~~~~
\begin{eqnarray} \label{background}
X_0=\tilde X_0: &\text{`background' quantities, GSF ignored}.
\end{eqnarray}
%~~~~~~~~~~~~~~~~~~~~~~~~~~~~~~~~~~~~~~~~~~~~~~~~~~~~~~~~~~~~~~~~

\subsection{Orbital setup and equation of motion}

In both analyses we consider a pointlike particle of mass $\mu(=\tilde\mu)$,
in a circular orbit around a Schwarzschild black hole with mass
$M(=\tilde M)\gg\mu$. The background Schwarzschild metric is denoted
$g_{\alpha\beta}^{(0)}(=\tilde g_{\alpha\beta}^{(0)})$. At the limit $\mu\to 0$
(i.e., ignoring GSF effects) the particle moves on a geodesic of
$g_{\alpha\beta}^{(0)}$, with a fixed Schwarzschild radius $r=r_0(=\tilde r_0)$.
(Note that identifying $\mu=\tilde\mu$, $M=\tilde M$ and
$r_0=\tilde r_0$ is sufficient to guarantee that the underlying physical
setups in both BS and SD are identical.) In both analyses we set up our
Schwarzschild coordinate system such that the background geodesic is
confined to the equatorial plane, $\theta=\pi/2$.

Now consider leading-order effects arising from the finite mass of the particle.
Denote the retarded linear metric perturbations due to the particle by $h_{\alpha\beta}$
and $\tilde h_{\alpha\beta}$ (each $\propto \mu$), corresponding to SD and BS.
The two perturbations are mathematically distinct, since $h_{\alpha\beta}$ is
taken in the Regge-Wheeler gauge \cite{Detweiler:2008}, whereas $\tilde h_{\alpha\beta}$
is taken in the Lorenz gauge \cite{Barack:2007tm}. In BS, the (regularized)
back-reaction from $\tilde h_{\alpha\beta}$ is seen as giving rise to a GSF
$\tilde F^{\alpha}$, which accelerates the particle on the background spacetime
$g_{\alpha\beta}^{(0)}$.
%At each instant, the true orbit is assumed to lie tangent
%to this reference background geodesic (see Ref.\ \cite{Pound&Poisson1} for an exact
%formulation of this assumption).
SD, on the other hand, does not calculate the GSF directly. Instead, he constructs
the smooth field $h^{\rm R}_{\alpha\beta}$ (`R' field), which is a particular vacuum
solution of the perturbed Einstein equations (again, in the Regge-Wheeler gauge),
with the property that the perturbed orbit is a geodesic of the total (smooth) perturbed
metric $g_{\alpha\beta}\equiv g_{\alpha\beta}^{(0)}+h^{\rm R}_{\alpha\beta}$.
SD then attributes back-reaction effects to the shift in the values of
the orbital parameters with respect to their unperturbed values.

%The Lorenz-gauge `R' field, to be denoted here $h^{\rm R}_{\alpha\beta}$, was
%not needed in the BS analysis, and has not been calculated in Ref.\
%\cite{Barack:2007tm}. However, we will need it in the current work for our
%comparison. The (numerical) construction of $h^{\rm R}_{\alpha\beta}$ will be
%described later.

%By ``regularizing''
%$\tilde h_{\alpha\beta}$ (see \cite{Detweiler:2008}), SD further constructs
%the `R' field, $\tilde h^{\rm R}_{\alpha\beta}$, which is a smooth vacuum
%solution of the perturbed Einstein equations (again, in the Regge-Wheeler gauge).
%BS do not calculate the corresponding Lorenz-Gauge R-field, $h^{\rm R}_{\alpha\beta}$.
%In BS, the (properly regularized) back-reaction from $h_{\alpha\beta}$ is seen
%as accelerating the particle with respect to an instantaneous geodesic of
%$g_{\alpha\beta}^0$. In contrast, SD describes the perturbed orbit of the
%particle as a geodesic of the total perturbed metric,
%$\tilde g_{\alpha\beta}\equiv g_{\alpha\beta}^0+\tilde h^{\rm R}_{\alpha\beta}$.

Let us parameterize the perturbed orbits in BS and SD by their respective proper
times $\tilde\tau$ and $\tau$. We need here two different symbols, because the two
orbits are formally defined in two different spacetimes:
$g_{\alpha\beta}^{(0)}$ and $g_{\alpha\beta}$, correspondingly. Let events along the
(accelerated) BS orbit have Schwarzschild coordinate values $\tilde x^{\alpha}_{\rm p}
(\tilde\tau)$, and events along the (geodesic) SD orbit have ``perturbed'' Schwarzschild
coordinate values $x^{\alpha}_{\rm p}(\tau)$ [such that, for a given physical point along the
orbit, $x^{\alpha}_{\rm p}-\tilde x^{\alpha}_{\rm p}=O(\mu)$].
Then define the corresponding four velocities $\tilde u^{\alpha}\equiv
d\tilde x_{\rm p}^{\alpha}/d\tilde\tau$ and $u^{\alpha}= dx_{\rm
p}^{\alpha}/d\tau$. (Note $\tilde u^{\alpha}$ and $u^{\alpha}$ are formally vectors
in two different geometries.) In our setup we have $u^\theta=\tilde u^{\theta}=0$,
and we impose the circularity conditions $u^r,\tilde u^{r}=0$ and $du^r/d\tau,\,
d\tilde u^{r}/d\tilde\tau=0$.\footnote{
These strict circularity conditions effectively ``switch off'' the dissipative
component of the GSF. Such an exercise enables us to study the instantaneous
conservative effect in isolation. Of course, a consistent treatment of the
long-term evolution would have to account for the dissipative orbital decay.
%but nonzero, radial velocity component, $u^r\propto O(\mu)$.
%Such an exercise is meaningful because the dissipative
%orbital decay in the systems of interest is very slow.
%A fully consistent treatment of the motion would, of course,
%account for the dissipative shrinkage of the orbital radius, allowing for a small,
%but nonzero, radial velocity component, $u^r\propto O(\mu)$. In situations where
%As explained in Ref.\ \cite{Detweiler:2008}, the condition $du^r/d\tau= O(\mu^2)$
%is imposed in order to distinguish our circular orbit from an orbit with
%a small eccentricity of order $\mu$---the latter having $u^r=O(\mu)$ as well as
%$du^r/d\tau=O(\mu)$. That this condition is consistent with the equation of motion
%at $O(\mu)$ is demonstrated by explicit construction---cf.\ Eqs.\
%(\ref{OmegaBS})--(\ref{utSD}) below.}
}
%In BS, tensorial
%indices are raised and lowered using the background metric, and, in particular,
%the normalization of the four-velocity along $\tilde x^{\alpha}_{\rm p}(\tilde\tau)$
%takes the form
%%~~~~~~~~~~~~~~~~~~~~~~~~~~~~~~~~~~~~~~~~~~~~~~~~~~~~~~~~~~~~~~~~
%\begin{equation} \label{unorm}
%g^0_{\alpha\beta}\tilde u^{\alpha}\tilde u^{\beta}=-1.
%\end{equation}
%%~~~~~~~~~~~~~~~~~~~~~~~~~~~~~~~~~~~~~~~~~~~~~~~~~~~~~~~~~~~~~~~~
%SD, on the other hand, raises and lowers indices using the (smooth) total
%metric $g_{\alpha\beta}$, and the four-velocity is normalized along
%$x^{\alpha}_{\rm p}(\tau)$ as
%%~~~~~~~~~~~~~~~~~~~~~~~~~~~~~~~~~~~~~~~~~~~~~~~~~~~~~~~~~~~~~~~~
%\begin{equation} \label{tildeunorm}
%g_{\alpha\beta}u^{\alpha}u^{\beta}=
%(g^0_{\alpha\beta}+h^{\rm R}_{\alpha\beta})
%u^{\alpha} u^{\beta}=-1.
%\end{equation}
%%~~~~~~~~~~~~~~~~~~~~~~~~~~~~~~~~~~~~~~~~~~~~~~~~~~~~~~~~~~~~~~~~
A given physical event along the orbit will generally have
$x^{\alpha}\ne\tilde x^{\alpha}$ and $u^{\alpha}\ne\tilde u^{\alpha}$,
%due both to the different reference
%geodesics and the different gauges employed in the two analysis;
with equalities restored only at the limit $\mu\to 0$. An explicit relation
between the corresponding tilded and non-tilded perturbed orbital quantities
will be established in the next section.

With the above notation, the equations of motion in BS and SD take the
respective forms
%~~~~~~~~~~~~~~~~~~~~~~~~~~~~~~~~~~~~~~~~~~~~~~~~~~~~~~~~~~~~~~~~
\begin{equation} \label{EOM:BS}
\frac{d\tilde u^{\alpha}}{d\tilde\tau}+\Gamma^{\alpha(0)}_{\beta\gamma}
\tilde u^{\beta}\tilde u^{\gamma}
=\mu^{-1}\tilde F^{\alpha}
\end{equation}
%~~~~~~~~~~~~~~~~~~~~~~~~~~~~~~~~~~~~~~~~~~~~~~~~~~~~~~~~~~~~~~~~
and
%~~~~~~~~~~~~~~~~~~~~~~~~~~~~~~~~~~~~~~~~~~~~~~~~~~~~~~~~~~~~~~~~
\begin{equation} \label{EOM:SD}
\frac{du^{\alpha}}{d\tau}
+\Gamma^{\alpha}_{\beta\gamma}u^{\beta}u^{\gamma}=0.
\end{equation}
%~~~~~~~~~~~~~~~~~~~~~~~~~~~~~~~~~~~~~~~~~~~~~~~~~~~~~~~~~~~~~~~~
Here the connection coefficients $\Gamma^{\alpha(0)}_{\beta\gamma}$ correspond to
the background metric $g^{(0)}_{\alpha\beta}$, and
$\Gamma^{\alpha}_{\beta\gamma}$ correspond to the
smooth perturbed metric $g_{\alpha\beta}$.
Note that in BS the GSF effect is entirely accounted for
by the acceleration term appearing on the right-hand side (RHS) of the equation of
motion, whereas in SD finite-$\mu$ effects are encoded in the
perturbed values of $u^{\alpha}$ and $\Gamma^{\alpha}_{\beta\gamma}$.
%Note also that the symmetry of the physical setup implies
%$u^\theta=\tilde u^{\theta}=0$, as well as $u^r,\tilde u^{r}\propto O(\mu)$
%and $du^r/d\tau,\, d\tilde u^{r}/d\tilde\tau\propto O(\mu^2)$.

\subsection{Gauge-invariant conservative GSF effects}

A gauge transformation in
$g_{\alpha\beta}^{(0)}+h^{\rm R}_{\alpha\beta}$ of the form
%~~~~~~~~~~~~~~~~~~~~~~~~~~~~~~~~~~~~~~~~~~~~~~~~~~~~~~~~~~~~~~~~
\begin{equation} \label{xi}
x^{\mu}\to x'^{\alpha}=x^{\alpha}+\xi^{\alpha},
\end{equation}
%~~~~~~~~~~~~~~~~~~~~~~~~~~~~~~~~~~~~~~~~~~~~~~~~~~~~~~~~~~~~~~~~
where $\xi^{\alpha}$ is of $O(\mu)$, will change the SD vacuum perturbation
$h^{\rm R}_{\alpha\beta}$
%retarded BS metric perturbations $\tilde h_{\alpha\beta}$
by an amount
%%~~~~~~~~~~~~~~~~~~~~~~~~~~~~~~~~~~~~~~~~~~~~~~~~~~~~~~~~~~~~~~~~
%\begin{equation} \label{h-gauge}
%\delta_{\xi}\tilde h_{\alpha\beta}
%=-(\xi_{\alpha;\beta}+\xi_{\beta;\alpha}),
%\end{equation}
%%~~~~~~~~~~~~~~~~~~~~~~~~~~~~~~~~~~~~~~~~~~~~~~~~~~~~~~~~~~~~~~~~
%where a semicolon denotes covariant differentiation with respect to
%$g^{(0)}_{\alpha\beta}$.
%[Note that differentiating instead with respect to
%$g_{\alpha\beta}$ would leave Eq.\ (\ref{h-gauge}) unaffected
%through $O(\mu)$, since $\xi^{\alpha}$ is already $O(\mu)$.]
%The SD vacuum perturbation $h^{\rm R}_{\alpha\beta}$
%will transforms in just the same way:
%(see Ref.\ \cite{Barack:2001ph}; this holds as long
%as long as the gauge transformation is sufficiently regular):
%~~~~~~~~~~~~~~~~~~~~~~~~~~~~~~~~~~~~~~~~~~~~~~~~~~~~~~~~~~~~~~~~
\begin{equation} \label{hR-gauge}
\delta_{\xi}h^{\rm R}_{\alpha\beta}
=-(\xi_{\alpha;\beta}+\xi_{\beta;\alpha}).
\end{equation}
%~~~~~~~~~~~~~~~~~~~~~~~~~~~~~~~~~~~~~~~~~~~~~~~~~~~~~~~~~~~~~~~~
%where a semicolon denotes covariant differentiation with respect to
%$g^{(0)}_{\alpha\beta}$.
Under a similar gauge transformation in
$g_{\alpha\beta}^{(0)}+\tilde h_{\alpha\beta}$,
the GSF in Eq.\ (\ref{EOM:BS}) will vary by an amount \cite{Barack:2001ph}
%~~~~~~~~~~~~~~~~~~~~~~~~~~~~~~~~~~~~~~~~~~~~~~~~~~~~~~~~~~~~~~~~
\begin{equation}\label{deltaF}
\delta_{\xi} \tilde F^{\alpha}=
\mu\left[\left(\delta^{\alpha}_{\lambda}+u_0^{\alpha}u_{0\lambda}\right)
\ddot\xi^{\lambda}
+R^{\alpha(0)}_{\;\mu\lambda\nu}u_{0}^{\mu}\xi^{\lambda}u_{0}^{\nu}\right],
\end{equation}
%~~~~~~~~~~~~~~~~~~~~~~~~~~~~~~~~~~~~~~~~~~~~~~~~~~~~~~~~~~~~~~~~
where an overdot denotes covariant differentiation with respect to $\tilde\tau$,
$u_0^{\alpha}$ is the four-velocity along the background geodesic,
and $R^{\alpha(0)}_{\;\mu\lambda\nu}$ is the background Riemann tensor, evaluated
at the particle.\footnote{
Equation (\ref{deltaF}) only makes sense when $\xi^{\alpha}$ and
$\ddot\xi^{\alpha}$ are well defined on the worldline.
It is not immediately clear how to define the GSF in gauges related to the
Lorenz gauge through transformations $\xi^{\alpha}$ which do not
satisfy these conditions. See Ref.\ \cite{Barack:2001ph} for a discussion
of this issue.}
The orbits $x^{\alpha}_{\rm p}(\tau)$
and $\tilde x^{\alpha}_{\rm p}(\tau)$ themselves will obviously change under
(\ref{xi}) as $\delta_{\xi} x^{\alpha}_{\rm p}=\delta_{\xi}\tilde x^{\alpha}
_{\rm p}=\xi^{\alpha}$, and, taking $\tau$ and $\tilde\tau$ derivatives,
respectively, we find for the corresponding four-velocities
%~~~~~~~~~~~~~~~~~~~~~~~~~~~~~~~~~~~~~~~~~~~~~~~~~~~~~~~~~~~~~~~~
\begin{equation} \label{u-gauge}
\delta_{\xi}u^{\alpha}=\delta_{\xi}\tilde u^{\alpha}=
\frac{d\xi^{\alpha}}{d\tau}.
\end{equation}
%~~~~~~~~~~~~~~~~~~~~~~~~~~~~~~~~~~~~~~~~~~~~~~~~~~~~~~~~~~~~~~~~
[Note that we may use $d\xi^{\alpha}/d\tau$ and $d\xi^{\alpha}/d\tilde\tau$
interchangeably, as the two quantities differ only by an amount of $O(\mu^2)$.]

What quantities can we construct, in our circular-equatorial orbit case,
which are independent of the gauge? Apart from the obvious $\mu$ and $\tau$,
two familiar gauge-invariant quantities are $-du_t/d\tau$
and $du_\varphi/d\tau$ (and their `tilded' counterparts)---the rate of change
of the geodesic energy and angular momentum parameters, respectively.
These quantities vanish at $\mu\to 0$, and at the leading order [$O(\mu)$]
are entirely due to the {\em dissipative} effect of the GSF.

For our BS/SD comparison we shall be interested in orbital gauge-invariants
which are affected by the {\em conservative} component of the GSF. Let us
restrict attention to a class of gauge transformations whose members,
denoted $\bar\xi^{\alpha}$, have $\bar\xi^{\theta}=0$ as well as the helical
symmetry
%~~~~~~~~~~~~~~~~~~~~~~~~~~~~~~~~~~~~~~~~~~~~~~~~~~~~~~~~~~~~~~~~
\begin{equation} \label{pr}
(\partial_t + \Omega_0\partial_\phi)\bar\xi^{\alpha}=0,
\end{equation}
%~~~~~~~~~~~~~~~~~~~~~~~~~~~~~~~~~~~~~~~~~~~~~~~~~~~~~~~~~~~~~~~~
where $\Omega_0=u^\varphi_0/u^t_0=\sqrt{M/r_0^3}$ is the orbital frequency of
the background geodesic. This choice is motivated as follows: If the original
metric perturbation respects the helical symmetry of the physical (black hole +
particle) configuration, and is also symmetric under reflection through the
equatorial plane---then a gauge transformation within the family $\bar\xi$
would retain these symmetries.
%(Indeed, these are not the {\em only} transformations which conform with spacetime
%symmetry, and we shall discuss an extension of the family $\bar\xi^{\alpha}$ in
%the next section.)
The condition (\ref{pr}) implies that, along the orbit,
$d\bar \xi^{\alpha}/d\tau=d\bar \xi^{\alpha}/d\tilde\tau=0$ through $O(\mu)$.
Equation (\ref{u-gauge}) therefore tells us that all components $u^{\alpha}$
(and $\tilde u^{\alpha}$) are invariant under $\bar \xi$.
Of these four components, only $u^{t}$ and $u^{\varphi}$ exhibit
non-trivial conservative $O(\mu)$ effects, and so could serve usefully
as gauge-invariants for our comparison. In fact, we shall follow here SD,
and utilize the alternative pair $U\equiv u^t$ and
%~~~~~~~~~~~~~~~~~~~~~~~~~~~~~~~~~~~~~~~~~~~~~~~~~~~~~~~~~~~~~~~~
\begin{equation} \label{GI}
\Omega\equiv u^{\varphi}/u^t =d\varphi_{\rm p}/dt_{\rm p},
\end{equation}
%~~~~~~~~~~~~~~~~~~~~~~~~~~~~~~~~~~~~~~~~~~~~~~~~~~~~~~~~~~~~~~~~
along with their `tilded' counterparts $\tilde U\equiv \tilde u^t$ and
$\tilde\Omega$.
The physical interpretation of the gauge invariant $U$ is somewhat
less obvious than that of the orbital frequency $\Omega$; it is discussed
in SD. Here we shall refer to $U$ (or $\tilde U$)
as the {\it time function} along the orbit.

The orbital frequency and time function are given in BS, through
$O(\mu)$, by
%~~~~~~~~~~~~~~~~~~~~~~~~~~~~~~~~~~~~~~~~~~~~~~~~~~~~~~~~~~~~~~~~
\begin{equation}\label{OmegaBS}
\tilde\Omega =
\Omega_0 \left[
1 - \frac{r_0(r_0-3M)}{2\mu M} \tilde F_r
\right],
\end{equation}
%~~~~~~~~~~~~~~~~~~~~~~~~~~~~~~~~~~~~~~~~~~~~~~~~~~~~~~~~~~~~~~~~
%~~~~~~~~~~~~~~~~~~~~~~~~~~~~~~~~~~~~~~~~~~~~~~~~~~~~~~~~~~~~~~~~
\begin{equation}\label{utBS}
\tilde U =
U_0\left(1- \frac{r_0}{2\mu} \tilde F_r
\right),
\end{equation}
%~~~~~~~~~~~~~~~~~~~~~~~~~~~~~~~~~~~~~~~~~~~~~~~~~~~~~~~~~~~~~~~~
where $U_0\equiv (1-3M/r_0)^{-1/2}$ is the time function along the
background geodesic. The corresponding SD quantities are given by
%~~~~~~~~~~~~~~~~~~~~~~~~~~~~~~~~~~~~~~~~~~~~~~~~~~~~~~~~~~~~~~~~
\begin{equation}\label{OmegaSD}
\Omega =
\Omega_0 \left[
1 - \frac{r_0(r_0-3M)}{4M}\,u^\alpha u^\beta h_{\alpha\beta,r}^{{\rm R}}
\right],
\end{equation}
%~~~~~~~~~~~~~~~~~~~~~~~~~~~~~~~~~~~~~~~~~~~~~~~~~~~~~~~~~~~~~~~~
%~~~~~~~~~~~~~~~~~~~~~~~~~~~~~~~~~~~~~~~~~~~~~~~~~~~~~~~~~~~~~~~~
\begin{equation}\label{utSD}
U=
U_0\left(
1
+ \frac{1}{2}u^\alpha u^\beta  h_{\alpha\beta}^{{\rm R}}
-\frac{r_0}{4}\,u^\alpha u^\beta h_{\alpha\beta,r}^{{\rm R}}
\right),
\end{equation}
%~~~~~~~~~~~~~~~~~~~~~~~~~~~~~~~~~~~~~~~~~~~~~~~~~~~~~~~~~~~~~~~~
where $h_{\alpha\beta}^{{\rm R}}$ and $h_{\alpha\beta,r}^{{\rm R}}$
are, of course, evaluated on the circular orbit.
That the quantities in Eqs.\ (\ref{OmegaBS})--(\ref{utSD}) are gauge
invariant under $\bar\xi$ can be readily verified with an explicit
calculation, using $\delta_{\bar\xi}r_0=\bar\xi^r$ and
%%~~~~~~~~~~~~~~~~~~~~~~~~~~~~~~~~~~~~~~~~~~~~~~~~~~~~~~~~~~~~~~~~
%\begin{equation}\label{deltaFr}
$\delta_{\bar\xi} \tilde F_{r}=-3\mu M\bar\xi^{r}/[r_0^2(r_0-3M)]$
%,\quad\quad
%%\delta F_{\theta}=\mu({\cal L}_0/r_0)^2\xi^{\theta}.
%\end{equation}
%%~~~~~~~~~~~~~~~~~~~~~~~~~~~~~~~~~~~~~~~~~~~~~~~~~~~~~~~~~~~~~~~~
(see Ref.\ \cite{Barack:2007tm}), along with Eq.\ (\ref{hR-gauge}).

%***********************************************************************
\section{Relation between BS and SD gauge invariants}\label{Sec:III}
%***********************************************************************

We now seek to obtain an explicit relation between $\Omega $ and $\tilde\Omega$
and between $U$ and $\tilde U$. We achieve this in two steps. In the first
step we map the BS trajectory onto a geodesic $\tilde y(\tau)$ defined \`{a} la
SD in a perturbed
spacetime (but in the Lorenz gauge). We define the quantities $\tilde\Omega_{y}$
and $\tilde U_{y}$ associated with this geodesic, and relate them
to the original $\tilde\Omega$ and $\tilde U$.
We then discuss the (somewhat unexpected) fact that the BS and
SD metric perturbations are related by a gauge transformation which is
{\em not} within the family $\bar\xi$. In the second step we therefore
work out a gauge transformation which brings the two perturbations within
a $\bar\xi$ transformation, and use this to establish an explicit relation
between $\tilde\Omega_{y}$, $\tilde U_{y}$ and the SD quantities
$\Omega$, $U$.

\subsection{Mapping between BS and SD trajectories}

In BS, the orbit $\tilde x_{\rm p}^{\alpha}(\tilde\tau)$ is an (accelerated)
trajectory in the Schwarzschild background $g_{\alpha\beta}^{(0)}$.
However, one can reinterpret this orbit, in the spirit of Detweiler and
Whiting \cite{Detweiler:2002mi}, as a geodesic of a smooth perturbed geometry
$\tilde g_{\alpha\beta}\equiv g_{\alpha\beta}^{(0)}+\tilde h_{\alpha\beta}^{\rm R}$, where
$\tilde h_{\alpha\beta}^{\rm R}$ is the R part of the BS (Lorenz-gauge)
metric perturbation $\tilde h_{\alpha\beta}$. (The field $\tilde h_{\alpha\beta}^{\rm R}$
has not been constructed explicitly in Ref.\ \cite{Barack:2007tm} as it is
not needed for calculating the GSF in the BS approach. This field can, in principle,
be constructed following the prescription of SD but working in the Lorenz gauge.)
This geodesic is physically identical with the SD geodesic $x_{\rm p}^{\alpha}(\tau)$,
since $h_{\alpha\beta}^{\rm R}$ and $\tilde h_{\alpha\beta}^{\rm R}$ represent the
same physical perturbation (merely expressed in different gauges).
We may therefore parameterize it  by the SD proper time $\tau$. Let us then denote
this geodesic by $\tilde y_{\rm p}^{\alpha}(\tau)$, with associated four-velocity
$\tilde v^{\alpha}(\tau)\equiv d\tilde y_{\rm p}^{\alpha}/d\tau$. Here the `tildes'
are meant to remind us that these quantities are described in the Lorenz gauge of BS,
but one should bear in mind that both $\tilde y_{\rm p}^{\alpha}$ and $\tilde v^{\alpha}$
are defined in the {\em perturbed} geometry $\tilde g_{\alpha\beta}$.

Our goal now will be to relate the vector $\tilde v^{\alpha}(\tau)$ to the BS
four-velocity $\tilde u^{\alpha}(\tilde\tau)$. This task is somewhat delicate,
because the two vectors are defined in two different spacetimes, and so one first
needs to establish a concrete mapping between the two trajectories
$\tilde y_{\rm p}^{\alpha}(\tau)$ and $\tilde x_{\rm p}^{\alpha}(\tilde\tau)$.
The mapping procedure (and its intimate relation with the gauge freedom of the
GSF) has been discussed in detail by Barack and Ori \cite{Barack:2001ph}.
The following derivation is largely inspired by that work.

We begin by obtaining a relation between the proper-time parameters $\tau$ and
$\tilde\tau$ for a given physical point along the orbit. The $\tau$-interval along
the geodesic $\tilde y_{\rm p}^{\alpha}(\tau)$ in $\tilde g_{\alpha\beta}$ satisfies
%~~~~~~~~~~~~~~~~~~~~~~~~~~~~~~~~~~~~~~~~~~~~~~~~~~~~~~~~~~~~~~~~
\begin{equation}\label{dtau2}
d\tau^2=-(g_{\alpha\beta}^{(0)}+\tilde h_{\alpha\beta}^{\rm R})
\, d\tilde y_{\rm p}^{\alpha}\, d\tilde y_{\rm p}^{\beta}.
\end{equation}
%~~~~~~~~~~~~~~~~~~~~~~~~~~~~~~~~~~~~~~~~~~~~~~~~~~~~~~~~~~~~~~~~
The four-velocity $\tilde v^{\alpha}(\tau)$ along this geodesic satisfies
%~~~~~~~~~~~~~~~~~~~~~~~~~~~~~~~~~~~~~~~~~~~~~~~~~~~~~~~~~~~~~~~~
\begin{equation} \label{EOM:v}
\frac{d\tilde v^{\alpha}}{d\tau}
+\tilde\Gamma^{\alpha}_{\beta\gamma}(\tilde y_{\rm p})\tilde v^{\beta}\tilde v^{\gamma}=0,
\end{equation}
%~~~~~~~~~~~~~~~~~~~~~~~~~~ ~~~~~~~~~~~~~~~~~~~~~~~~~~~~~~~~~~~~~~
where the connections $\tilde\Gamma^{\alpha}_{\beta\gamma}$ are those associated
with the full metric $\tilde g_{\alpha\beta}$, and are here evaluated at
$\tilde y^{\alpha}_{\rm p}(\tau)$. We now wish to think of $g_{\alpha\beta}^{(0)}$ and
$\tilde h_{\alpha\beta}^{\rm R}$ as separate tensor fields in the geometry
$\tilde g_{\alpha\beta}$. In general, of course, the splitting of
$\tilde g_{\alpha\beta}$ into a ``background'' field $g_{\alpha\beta}^{(0)}$ and a
``perturbation'' field $\tilde h_{\alpha\beta}^{\rm R}$ depends on the choice of
gauge (indeed, it is the {\em origin} of the gauge freedom). Here, however, we
fix both the background coordinates (Schwarzschild) and the gauge (Lorenz), and
so both $g_{\alpha\beta}^{(0)}$ and $\tilde h_{\alpha\beta}^{\rm R}$ are defined
unambiguously. Let $\tilde\Gamma^{\alpha(0)}_{\beta\gamma}$ be the connections
associated with the field $g_{\alpha\beta}^{(0)}$, and let
$\Delta\tilde\Gamma^{\alpha}_{\beta\gamma}\equiv
\tilde\Gamma^{\alpha}_{\beta\gamma}-\tilde\Gamma^{\alpha(0)}_{\beta\gamma}$.
Then Eq.\ (\ref{EOM:v}) can be written in the form
%~~~~~~~~~~~~~~~~~~~~~~~~~~~~~~~~~~~~~~~~~~~~~~~~~~~~~~~~~~~~~~~~
\begin{equation} \label{EOM:v2}
\frac{d\tilde v^{\alpha}}{d\tau}
+\tilde\Gamma^{\alpha(0)}_{\beta\gamma}(\tilde y_{\rm p})\tilde v^{\beta}\tilde v^{\gamma}
=-\Delta\tilde\Gamma^{\alpha}_{\beta\gamma}(\tilde y_{\rm p})\tilde v^{\beta}\tilde v^{\gamma}.
\end{equation}
%~~~~~~~~~~~~~~~~~~~~~~~~~~ ~~~~~~~~~~~~~~~~~~~~~~~~~~~~~~~~~~~~~~

Next, {\em define} a new parameter $\tilde\tau$ along the geodesic
$\tilde y_{\rm p}^{\alpha}(\tau)$ using
%~~~~~~~~~~~~~~~~~~~~~~~~~~~~~~~~~~~~~~~~~~~~~~~~~~~~~~~~~~~~~~~~
\begin{equation}\label{dtildetau2}
d\tilde\tau^2=-g_{\alpha\beta}^{(0)}\, d\tilde y_{\rm p}^{\alpha}\,
d\tilde y_{\rm p}^{\beta}
\end{equation}
%~~~~~~~~~~~~~~~~~~~~~~~~~~~~~~~~~~~~~~~~~~~~~~~~~~~~~~~~~~~~~~~~
[with the requirement that the ``zero point'' of $\tilde\tau$ is chosen
such that $\tilde\tau-\tau=O(\mu)$]; and {\em define} the new tangent
vector
%~~~~~~~~~~~~~~~~~~~~~~~~~~~~~~~~~~~~~~~~~~~~~~~~~~~~~~~~~~~~~~~~
\begin{equation}\label{tildeu}
\tilde u^{\alpha}= \frac{d\tilde y_{\rm p}^{\alpha}}{d\tilde\tau}=
\tilde v^{\alpha}\, \frac{d\tau}{d\tilde\tau}.
\end{equation}
%~~~~~~~~~~~~~~~~~~~~~~~~~~~~~~~~~~~~~~~~~~~~~~~~~~~~~~~~~~~~~~~~
We use here the symbols $\tilde\tau$ and $\tilde u^{\alpha}$ (in a
slight abuse of notation) as we are soon to interpret these as the BS proper
time and four velocity. For now, however, recall that $\tilde\tau$ and
$\tilde u^{\alpha}$ are defined along the geodesic $\tilde y_{\rm p}^{\alpha}$ in the
{\em perturbed} spacetime $\tilde g_{\alpha\beta}$ (not in $g^{(0)}_{\alpha\beta}$);
but note $\tilde\tau$ and $\tilde u^{\alpha}$ are {\em not} proper time
and four-velocity along this geodesic.
Substituting $\tilde v^{\alpha}=(d\tilde\tau/d\tau)\tilde u^{\alpha}$
and $d/d\tau=(d\tilde\tau/d\tau)d/d\tilde\tau$ in Eq.\ (\ref{EOM:v2}),
we now obtain
%~~~~~~~~~~~~~~~~~~~~~~~~~~~~~~~~~~~~~~~~~~~~~~~~~~~~~~~~~~~~~~~~
\begin{equation} \label{EOM:v3}
\frac{d\tilde u^{\alpha}}{d\tilde\tau}
+\tilde\Gamma^{\alpha(0)}_{\beta\gamma}(\tilde y_{\rm p})
\tilde u^{\beta}\tilde u^{\gamma}
=-\Delta\tilde\Gamma^{\alpha}_{\beta\gamma}(\tilde y_{\rm p})
\tilde u^{\beta}\tilde u^{\gamma}-\beta \tilde u^{\alpha},
\end{equation}
%~~~~~~~~~~~~~~~~~~~~~~~~~~ ~~~~~~~~~~~~~~~~~~~~~~~~~~~~~~~~~~~~~~
where $\beta\equiv(d\tau/d\tilde\tau)^2(d^2\tilde\tau/d\tau^2)$.
The expression on the left-hand side here can be {\em interpreted}
as the acceleration of a trajectory $\tilde y_{\rm p}(\tilde\tau)$
(with proper time $\tilde\tau$ and four-velocity $\tilde u^{\alpha}$)
in a spacetime $g^{(0)}_{\alpha\beta}$. This acceleration is orthogonal
to $\tilde u^{\alpha}$ in the spacetime $g^{(0)}_{\alpha\beta}$
[by virtue of $g^{(0)}_{\alpha\beta}\tilde u^{\alpha}\tilde u^{\beta}=-1$,
which follows from Eq.\ (\ref{dtildetau2})], so formally projecting both
sides of Eq.\ (\ref{EOM:v3}) orthogonally to $\tilde u^{\alpha}$
(in the metric $g^{(0)}_{\alpha\beta}$) finally gives
%~~~~~~~~~~~~~~~~~~~~~~~~~~~~~~~~~~~~~~~~~~~~~~~~~~~~~~~~~~~~~~~~
\begin{equation} \label{EOM:v4}
\frac{d\tilde u^{\alpha}}{d\tilde\tau}
+\tilde\Gamma^{\alpha(0)}_{\beta\gamma}(\tilde y_{\rm p})
\tilde u^{\beta}\tilde u^{\gamma}
=-(\delta^{\alpha}_{\mu}+\tilde u^{\alpha}\tilde u^{\nu}g^{(0)}_{\nu\mu})
\Delta\tilde\Gamma^{\mu}_{\beta\gamma}(\tilde y_{\rm p})
\tilde u^{\beta}\tilde u^{\gamma}.
\end{equation}
%~~~~~~~~~~~~~~~~~~~~~~~~~~ ~~~~~~~~~~~~~~~~~~~~~~~~~~~~~~~~~~~~~~

Now, Detweiler and Whiting have shown \cite{Detweiler:2002mi} that
the expression on the RHS of Eq.\ (\ref{EOM:v4}) (where, recall,
$\Delta\tilde\Gamma^{\mu}_{\beta\gamma}$ is derived from the
R-field $\tilde h_{\alpha\beta}^{\rm R}$) {\em is precisely the
self-acceleration} $\mu^{-1}\tilde F^{\alpha}$, if one interprets this quantity
as a vector in $g^{(0)}_{\alpha\beta}$. Comparing then the forms of Eqs.\
(\ref{EOM:v4}) and (\ref{EOM:BS}), we arrive at the following conclusion:
If each point along the geodesic $\tilde y_{\rm p}(\tau)$ is associated with
a point along the BS trajectory having {\em the same coordinate value} (i.e.\
$\tilde x_{\rm p}=\tilde y_{\rm p}$), then the quantities $\tilde\tau$ and
$\tilde u^{\alpha}$ defined in the perturbed spacetime $\tilde g^{\alpha\beta}$
in Eqs.\ (\ref{dtildetau2}) and (\ref{tildeu})
can be interpreted---and would have the same values---as the corresponding
BS quantities $\tilde\tau$ and $\tilde u^{\alpha}=d\tilde x_{\rm p}^{\alpha}
/d\tilde\tau$.

The main practical outcome from the above discussion are formulas relating
the BS quantities $\tilde\tau$ and $\tilde u^{\alpha}$ to their counterparts
$\tau$ and $\tilde v^{\alpha}$ defined \`{a} la SD in the perturbed spacetime.
From Eqs.\ (\ref{dtau2}) and (\ref{dtildetau2}) we obtain
%~~~~~~~~~~~~~~~~~~~~~~~~~~~~~~~~~~~~~~~~~~~~~~~~~~~~~~~~~~~~~~~~
\begin{equation}\label{tauratio}
\frac{d\tilde\tau}{d\tau}=1+\tilde H
\end{equation}
%~~~~~~~~~~~~~~~~~~~~~~~~~~~~~~~~~~~~~~~~~~~~~~~~~~~~~~~~~~~~~~~~
[through $O(\mu)$], where
%~~~~~~~~~~~~~~~~~~~~~~~~~~~~~~~~~~~~~~~~~~~~~~~~~~~~~~~~~~~~~~~~
\begin{equation}\label{H}
\tilde H\equiv \frac{1}{2}\tilde h_{\alpha\beta}^{{\rm R}}(x_{\rm p})
\tilde u^\alpha \tilde u^\beta.
%=\tilde h_{\alpha\beta}^{{\rm R}}u^\alpha u^\beta
%\quad \text{(gauge invariant)}.
\end{equation}
%~~~~~~~~~~~~~~~~~~~~~~~~~~~~~~~~~~~~~~~~~~~~~~~~~~~~~~~~~~~~~~~~
Equation (\ref{tauratio}) describes the relation between the BS and SD
proper-time parameters, assuming the ``same-coordinate-value'' mapping
between the respective trajectories. The relation between the
four-velocities then follows immediately from Eq.\ (\ref{tildeu}):
%~~~~~~~~~~~~~~~~~~~~~~~~~~~~~~~~~~~~~~~~~~~~~~~~~~~~~~~~~~~~~~~~
\begin{equation}\label{uBS}
\tilde v^{\alpha}=\left(1+\tilde H\right) \tilde u^{\alpha}
\end{equation}
%~~~~~~~~~~~~~~~~~~~~~~~~~~~~~~~~~~~~~~~~~~~~~~~~~~~~~~~~~~~~~~~~
[again, through $O(\mu)$].

Let us finally define the frequency
$\tilde\Omega_y\equiv \tilde v^{\varphi}/\tilde v^{t}$
and time function
$\tilde U_y\equiv \tilde v^{t}$ associated with the
trajectory $\tilde y^{\alpha}(\tau)$, and relate these to the BS
quantities $\tilde\Omega$ and $\tilde U$.
For the frequency we have, using Eq.\ (\ref{uBS}),
%$\tilde\Omega_y=\tilde u^{\varphi}/\tilde u^{t}$, and so
%~~~~~~~~~~~~~~~~~~~~~~~~~~~~~~~~~~~~~~~~~~~~~~~~~~~~~~~~~~~~~~~~
\begin{equation}\label{tildeOmegaBS}
\tilde\Omega_y =\tilde v^{\varphi}/\tilde v^{t}=
\tilde u^{\varphi}/\tilde u^{t}=\tilde\Omega.
%=\Omega_0 \left[1 - \frac{r_0(r_0-3M)}{2\mu M} \tilde F_r
%\right],
\end{equation}
%~~~~~~~~~~~~~~~~~~~~~~~~~~~~~~~~~~~~~~~~~~~~~~~~~~~~~~~~~~~~~~~~
%where the second equality has been copied over from Eq.\ (\ref{OmegaBS}).
For the time function, Eq.\ (\ref{uBS}) immediately gives
%$\tilde u_y^{t}=\left(1+H\right) \tilde u^{t}$, and so, using Eq.\
%(\ref{utBS}),
%~~~~~~~~~~~~~~~~~~~~~~~~~~~~~~~~~~~~~~~~~~~~~~~~~~~~~~~~~~~~~~~~
\begin{equation}\label{tildeutBS}
\tilde U_y=\left(1+\tilde H\right) \tilde U.
%\tilde u_y^t =
%u^t_0\left(1+H -\frac{r_0}{2\mu} \tilde F_r
%\right),
\end{equation}
%~~~~~~~~~~~~~~~~~~~~~~~~~~~~~~~~~~~~~~~~~~~~~~~~~~~~~~~~~~~~~~~~
%through $O(\mu)$.

\subsection{Adjusting the gauge}

If the BS perturbation $\tilde h_{\alpha\beta}^{{\rm R}}$ and the SD perturbation
$h_{\alpha\beta}^{{\rm R}}$ were related through a gauge transformation
within the family $\bar\xi$, then the quantities $\tilde \Omega_{y}$ and
$\tilde U_{y}$ would have to be equal to their SD counterparts
$\Omega$ and $U$, since, recall, $\Omega$ and $U$ are
gauge invariant under $\bar\xi$. As we now discuss, it turns out that
this is not the case: The two perturbations differ by a gauge transformation
{\em not} belonging to the family $\bar\xi$.

To see this, it suffices to examine the asymptotic form of the $tt$ components
of the BS and SD metric perturbations at $r\to\infty$.
In SD a gauge is chosen (within the class of Regge-Wheeler gauges) such that
the perturbed metric is asymptotically Minkowskian:
%~~~~~~~~~~~~~~~~~~~~~~~~~~~~~~~~~~~~~~~~~~~~~~~~~~~~~~~~~~~~~~~~
\begin{equation}\label{gttSD}
g_{tt}^{(0)} + h_{tt}^{{\rm R}} \rightarrow -1 \quad (r\rightarrow \infty).
\end{equation}
%~~~~~~~~~~~~~~~~~~~~~~~~~~~~~~~~~~~~~~~~~~~~~~~~~~~~~~~~~~~~~~~~
This is impossible to achieve with the BS Lorenz-gauge perturbation, which has
a monople contribution that fails to vanish at $r\rightarrow\infty$. This monopole
contribution was first derived by Detweiler and Poisson \cite{Detweiler:2003ci},
who showed that it {\em uniquely} describes the correct mass perturbation due to
the particle: Any other Lorenz-gauge monopole solution would either diverge at
the event horizon or at infinity, and/or fail to describe the correct mass perturbation.
The field $\tilde h_{\alpha\beta}^{{\rm R}}$ inherits the large-$r$ asymptotic form
of the full perturbation $\tilde h_{\alpha\beta}$, which is dominated by the above
monopole term. The perturbed BS metric has the asymptotic form
%~~~~~~~~~~~~~~~~~~~~~~~~~~~~~~~~~~~~~~~~~~~~~~~~~~~~~~~~~~~~~~~~
\begin{equation}\label{gttBS}
 g^{(0)}_{tt} + \tilde h_{tt}^{{\rm R}} \rightarrow -1-2\alpha
\quad (r\rightarrow \infty),
\end{equation}
%~~~~~~~~~~~~~~~~~~~~~~~~~~~~~~~~~~~~~~~~~~~~~~~~~~~~~~~~~~~~~~~~
where
%~~~~~~~~~~~~~~~~~~~~~~~~~~~~~~~~~~~~~~~~~~~~~~~~~~~~~~~~~~~~~~~~
\begin{equation}\label{alpha}
\alpha=\frac{\mu}{\sqrt{r_0(r_0-3M)}}\, ,
\end{equation}
%~~~~~~~~~~~~~~~~~~~~~~~~~~~~~~~~~~~~~~~~~~~~~~~~~~~~~~~~~~~~~~~~
and where the term $-2\alpha$ comes entirely from the monopole perturbation.
(The other tensorial components of the BS metric attain their Minkowski
values at $r\to\infty$, just like the SD metric.)
The difference between the asymptotic forms in Eqs.\ (\ref{gttSD}) and
(\ref{gttBS}) must be accounted for by a monopole gauge transformation.
From Eq.\ (\ref{hR-gauge}) we find that the generator of this gauge
transformation must satisfy, asymptotically, $\xi_{t,t}=\alpha$
(as well as $\xi_{t,\varphi}=0$, since a monople transformation cannot
depend on $\varphi$), which necessarily violates the condition of
Eq.\ (\ref{pr}). Hence, the BS and SD perturbations differ by a gauge
transformation {\em outside} the family $\bar\xi$.

The above conclusion by no means suggests that either of BS/SD's
metric perturbations violates the helical symmetry of the
physical spacetime (they both respect it, in fact). Rather, it calls for a more
careful inspection of the class of gauge transformations which maintain
helical symmetry. Consider a gauge transformation
$\xi^\alpha=a t\delta_t^{\alpha}$, where $a(\propto\mu)$ is a constant.
This transformation affects only the $tt$ component of the metric
perturbation, which shifts by an amount
$\delta_{\xi}h^{\rm R}_{tt}=2(1-2M/r)a$.
If the original perturbation is helically symmetric [i.e., satisfying
$(\partial_t + \Omega_0\partial_\phi)h^{\rm R}_{\alpha\beta}=0$],
then so will be the gauge-transformed perturbation, even though the
generator $\xi^{\alpha}$ itself does not respect the helical symmetry of
the geometry. This suggests a more general class of ``physically reasonable''
gauge transformations, with generators given by
%~~~~~~~~~~~~~~~~~~~~~~~~~~~~~~~~~~~~~~~~~~~~~~~~~~~~~~~~~~~~~~~~
\begin{equation}\label{hatxi}
\hat \xi^{\alpha}=\bar\xi^{\alpha}+ a\, t\, \delta_t^{\alpha},
\end{equation}
%~~~~~~~~~~~~~~~~~~~~~~~~~~~~~~~~~~~~~~~~~~~~~~~~~~~~~~~~~~~~~~~~
where $\bar\xi^{\alpha}$ is any vector satisfying Eq.\ (\ref{pr}), and
$a$ is any constant ($\propto\mu$). While the generators $\hat \xi$ themselves
are generally not helically symmetric, they do not interfere with the
helical symmetry of the metric perturbation they act upon. It is not
difficult to convince oneself that the class $\hat \xi$ is the most
general class with this property.\footnote{The class $\hat \xi^{\alpha}$ has
been introduced before in the literature---see Appendix A of Ref.\
\cite{Detweiler:2003ci}.}

The gauge transformation between the BS and SD perturbations belongs
to the class $\hat\xi$, with $a=\alpha\ne 0$. Crucially for our analysis,
the two quantities $\Omega$ and $U$, which are invariant under $\bar\xi$,
are not invariant under $\hat\xi$ (for any $a\ne 0$). More precisely,
we have (for $a=\alpha$)
%~~~~~~~~~~~~~~~~~~~~~~~~~~~~~~~~~~~~~~~~~~~~~~~~~~~~~~~~~~~~~~~~
\begin{equation}\label{deltaxi}
\delta_{\hat\xi} \Omega =-\alpha\, \Omega_0,  \quad\quad
\delta_{\hat\xi} U =\alpha\, U_0,
\end{equation}
%~~~~~~~~~~~~~~~~~~~~~~~~~~~~~~~~~~~~~~~~~~~~~~~~~~~~~~~~~~~~~~~~
through $O(\mu)$. These expressions are easily derived using Eq.\
(\ref{u-gauge}), noting $d\bar\xi^{\alpha}/d\tau=0$ and
hence $d\hat\xi^{\alpha}/d\tau=\alpha\,U\delta^{\alpha}_t$
along the worldline.
%\begin{eqnarray}
%\delta_{\xi} \left( u^\mu u^\nu h_{\mu\nu} \right) &=&
%\frac{2\alpha (r_0-2M)}{r_0-3M}, \\
%\delta_{\xi} F_r &=&
%\frac{2 \alpha \mu M}{r_0(r_0-3M)}, \\
%\delta_{\xi} \Omega^2 &=&
%-\frac{2M\alpha}{r_0^3}, \\
%\delta_{\xi} \tilde{u}^t &=&
%\alpha\sqrt{\frac{r_0}{r_0-3M}}
%\end{eqnarray}
Applying the gauge transformation $\hat\xi^{\alpha}=\alpha\, t\, \delta_t^{\alpha}$,
the quantities $\tilde \Omega_{y}$ and $\tilde U_{y}$ of
Eqs.\ (\ref{tildeOmegaBS}) and (\ref{tildeutBS}) transform as
%~~~~~~~~~~~~~~~~~~~~~~~~~~~~~~~~~~~~~~~~~~~~~~~~~~~~~~~~~~~~~~~~
\begin{eqnarray} \label{trans1}
\tilde \Omega_{y} &\rightarrow&
\tilde \Omega_{y}-\alpha\Omega_0=\tilde\Omega-\alpha\Omega_0
\nonumber \\ &&=
\tilde\Omega(1-\alpha) +O(\mu^2),
\end{eqnarray}
%~~~~~~~~~~~~~~~~~~~~~~~~~~~~~~~~~~~~~~~~~~~~~~~~~~~~~~~~~~~~~~~~
and
%~~~~~~~~~~~~~~~~~~~~~~~~~~~~~~~~~~~~~~~~~~~~~~~~~~~~~~~~~~~~~~~~
\begin{eqnarray} \label{trans2}
\tilde U_{y} &\rightarrow&
\tilde U_{y}+\alpha\, U_0=(1+\tilde H)\tilde U+\alpha\, U_0
\nonumber \\ &&=
\tilde U(1+\alpha+\tilde H) +O(\mu^2).
\end{eqnarray}
%~~~~~~~~~~~~~~~~~~~~~~~~~~~~~~~~~~~~~~~~~~~~~~~~~~~~~~~~~~~~~~~~
Since this gauge transformation brings the BS and SD perturbations to
within a $\bar\xi$ transformation from one other, the RHS
expressions in Eqs.\ (\ref{trans1}) and (\ref{trans2}) must be equal
to the SD quantities $\Omega$ and $U$, respectively
(as, recall, the frequency and time function are invariant under $\bar\xi$
transformations).
We hence arrive at the desired relations
%~~~~~~~~~~~~~~~~~~~~~~~~~~~~~~~~~~~~~~~~~~~~~~~~~~~~~~~~~~~~~~~~
\begin{equation} \label{relation1}
\Omega=\tilde\Omega(1-\alpha),
\end{equation}
%~~~~~~~~~~~~~~~~~~~~~~~~~~~~~~~~~~~~~~~~~~~~~~~~~~~~~~~~~~~~~~~~
%~~~~~~~~~~~~~~~~~~~~~~~~~~~~~~~~~~~~~~~~~~~~~~~~~~~~~~~~~~~~~~~~
\begin{equation} \label{relation2}
U =\tilde U\left(1 +\alpha + \tilde H \right),
\end{equation}
%~~~~~~~~~~~~~~~~~~~~~~~~~~~~~~~~~~~~~~~~~~~~~~~~~~~~~~~~~~~~~~~~
holding through $O(\mu)$.

Equations (\ref{relation1}) and (\ref{relation2}) explicitly relate
between the BS and SD values of the perturbed frequency and time function.
These relations do not involve the GSF, but they do require knowledge
of the quantity $\tilde H\equiv \tilde h_{\alpha\beta}^{{\rm R}}
\tilde u^\alpha \tilde u^\beta/2$,
which is to be constructed from the BS perturbation. A physical interpretation
of the quantity $\tilde H$ is suggested from Eq.\ (\ref{tauratio}): It
describes how proper-time intervals relate to each other in the different
orbital representations of BS and SD.
It is straightforward to check that $\tilde H$ is
invariant, for circular orbits, within the class of gauge transformations
$\bar\xi$ (though not within $\hat\xi$ for $a\ne 0$).

%***********************************************************************
\section{Comparison of numerical results}\label{Sec:IV}
%***********************************************************************

\subsection{Gauge-invariant comparison formula}

The relations (\ref{relation1}) and (\ref{relation2}), as they stand, do
not quite yet offer a practical means by which to test the BS/SD numerical results
against each other. The reason is as follows: The equalities expressed in
these relations hold, through $O(\mu)$, for a given physical orbit with
(perturbed)  SD radius $r_{\rm p}$ and BS radius $\tilde r_{\rm p}$.
In these equalities, $\Omega$ and $U$ are to be evaluated at $r_{\rm p}$,
while $\tilde\Omega$ and $\tilde U$ are to be evaluated at
$\tilde r_{\rm p}$. Alas, the relation between $r_{\rm p}$ and
$\tilde r_{\rm p}$ is not known to us at $O(\mu)$: It depends on the
precise gauge transformation between the BS and SD perturbations, which
we have not solved for. Without knowledge of how $r_{\rm p}$ relates to
$\tilde r_{\rm p}$ at $O(\mu)$, it is not possible to extract the $O(\mu)$ parts
of Eqs.\ (\ref{relation1}) and (\ref{relation2}).
Stated differently, we have the following problem: While the quantities
$\Omega$ and $U$ are gauge-invariant (under $\bar\xi$), the finite-mass
differences $\Omega-\Omega_0$ and $U-U_0$ (which are the quantities whose
numerical values we wish to test) are {\em not} gauge invariant, since
$\Omega_0$ and $U_0$ depend on the gauge (through $r_0$). We should
instead be looking at finite-$\mu$ corrections which are gauge
independent.

A standard solution to this problem is to express one gauge invariant in
terms of the other, and, following SD, we shall pursue this direction here.
SD introduced the gauge-invariant radius
%~~~~~~~~~~~~~~~~~~~~~~~~~~~~~~~~~~~~~~~~~~~~~~~~~~~~~~~~~~~~~~~~
\begin{equation}\label{R}
R\equiv \left( M/\Omega^2 \right)^{1/3}
%=
%r_0 \left[
%1 + \frac{2}{3}\alpha
%+ \frac{r_0(r_0-3M)}{3\mu M}F_r
%\right].
\end{equation}
%~~~~~~~~~~~~~~~~~~~~~~~~~~~~~ ~~~~~~~~~~~~~~~~~~~~~~~~~~~~~~~~~~~
(denoted $R_{\Omega}$ in Ref.\ \cite{Detweiler:2008}), and then expressed
$U$ in terms of $R$. The difference
%~~~~~~~~~~~~~~~~~~~~~~~~~~~~~~~~~~~~~~~~~~~~~~~~~~~~~~~~~~~~~~~~
\begin{equation}\label{Deltaut}
\Delta U(R) \equiv U(R) -(1-3M/R)^{-1/2}
\end{equation}
%~~~~~~~~~~~~~~~~~~~~~~~~~~~~~~~~~~~~~~~~~~~~~~~~~~~~~~~~~~~~~~~~
[which is of $O(\mu)$, since $(1-3M/R)^{-1/2}=U_0+O(\mu)$]
is then a genuinely gauge-invariant measure of the conservative finite-$\mu$
effect. SD derived the relation $\Delta U(R)$ numerically, and
utilized it for comparison with results from PN theory. Here we shall
reconstruct the relation $\Delta U(R)$ from BS quantities,
and use it to compare with SD.

Using Eq.\ (\ref{utBS}) to substitute for $\tilde U$ in Eq.\ (\ref{relation2}),
and then substituting the result for $U$ in Eq.\ (\ref{Deltaut}),
%in Eqs.\ (\ref{Deltaut}), (\ref{relation2}) and (\ref{utBS})
we obtain
%~~~~~~~~~~~~~~~~~~~~~~~~~~~~~~~~~~~~~~~~~~~~~~~~~~~~~~~~~~~~~~~~
\begin{equation}\label{Deltaut2}
\Delta U =(1-3M/r_0)^{-1/2}\left(1 +\alpha + \tilde H
-\frac{r_0}{2\mu}\tilde F_r \right)-(1-3M/R)^{-1/2}
+O(\mu^2).
\end{equation}
%~~~~~~~~~~~~~~~~~~~~~~~~~~~~~~~~~~~~~~~~~~~~~~~~~~~~~~~~~~~~~~~~
Our goal is to express the RHS here entirely in terms of the
gauge-invariant radius $R$. For this, we need first to express
$r_0$ in terms of $R$ up through $O(\mu)$. Starting from the
definition (\ref{R}), then using Eqs.\ (\ref{relation1}) and (\ref{OmegaBS})
in succession, and finally substituting $\Omega_0^{-2}=r_0^3/M$, we obtain
%~~~~~~~~~~~~~~~~~~~~~~~~~~~~~~~~~~~~~~~~~~~~~~~~~~~~~~~~~~~~~~~~
\begin{eqnarray}\label{R3r}
R^3&=&M\Omega^{-2}=M\tilde\Omega^{-2}(1-\alpha)^{-2}
\nonumber\\
&=& r_0^3 \left[1 +2\alpha+ \frac{r_0(r_0-3M)}{\mu M}\tilde F_r\right]
+O(\mu^2).
\end{eqnarray}
%~~~~~~~~~~~~~~~~~~~~~~~~~~~~~~~~~~~~~~~~~~~~~~~~~~~~~~~~~~~~~~~~
Hence,
%~~~~~~~~~~~~~~~~~~~~~~~~~~~~~~~~~~~~~~~~~~~~~~~~~~~~~~~~~~~~~~~~
\begin{equation}\label{r0}
r_0=R \left[1 -\frac{2}{3}\alpha+ \frac{R(R-3M)}
{3\mu M} \tilde F_r\right]+O(\mu^2),
\end{equation}
%~~~~~~~~~~~~~~~~~~~~~~~~~~~~~~~~~~~~~~~~~~~~~~~~~~~~~~~~~~~~~~~~
where the quantities $\alpha$ and $\tilde F_r/\mu$, which are already $O(\mu)$,
are evaluated at $R$.
%(so we henceforth take $\alpha=\mu[\tilde R(\tilde R-3M)]^{-1/2}$).
Substituting for $r_0(R)$ in Eq.\ (\ref{Deltaut2}) and expanding through
$O(\mu)$, we obtain
%~~~~~~~~~~~~~~~~~~~~~~~~~~~~~~~~~~~~~~~~~~~~~~~~~~~~~~~~~~~~~~~~
\begin{equation}\label{almostfinal}
\Delta U =(1-3M/R)^{-1/2}\left(
\frac{R-2M}{R-3M}\, \alpha + \tilde H \right) +O(\mu^2).
\end{equation}
%~~~~~~~~~~~~~~~~~~~~~~~~~~~~~~~~~~~~~~~~~~~~~~~~~~~~~~~~~~~~~~~~
Finally, we note that on the RHS we can replace $R\to r_0$ without
affecting $\Delta U$ at leading order [since $r_0-R\sim
O(\mu)$, and $\alpha$ and $\tilde H$ are already $O(\mu)$]. This allows us to
recast Eq.\ (\ref{almostfinal}) in a more practical form:
%~~~~~~~~~~~~~~~~~~~~~~~~~~~~~~~~~~~~~~~~~~~~~~~~~~~~~~~~~~~~~~~~
\begin{equation}\label{final}
\Delta U =(1-3M/r_0)^{-1/2}\left(
\frac{r_0-2M}{r_0-3M}\, \alpha + \tilde H \right),
\end{equation}
%~~~~~~~~~~~~~~~~~~~~~~~~~~~~~~~~~~~~~~~~~~~~~~~~~~~~~~~~~~~~~~~~
with corrections of $O(\mu^2)$.
Recall $\alpha=\mu[r_0(r_0-3M)]^{-1/2}$, and $\tilde H$ is the
quantity constructed from the BS metric perturbation through Eq.\ (\ref{H}).

It is interesting to point out that our final expression, Eq.\ (\ref{final}), bears
no direct reference to the GSF. This suggests that the $O(\mu)$ coordinate change
$r\to R$, in fact, amounts to a gauge transformation which sets the conservative
piece of the BS GSF to zero.
This is further suggested by noticing $U_0(r_0(R))=\tilde U(r_0\to R)$ (up to terms
$\propto\alpha$)---which one can readily verify starting with
$U_0= (1-3M/r_0)^{-1/2}$ and using Eqs.\ (\ref{r0}) and (\ref{utBS}).

\subsection{Numerical results}

We have reconstructed the relation $\Delta U(r_0)$ numerically, using the
BS Lorenz-gauge code, based on Eq.\ (\ref{final}).
The calculation of the full (retarded) Lorenz-gauge
perturbation is describe in detail in Ref.\ \cite{Barack:2007tm}. In short,
the method relies on the perturbation formalism of Barack and Lousto
\cite{Barack:2005nr}, in which the Lorenz-gauge perturbation equations are
decomposed into tensor harmonics, resulting in a set of 10 partial differential
equations (in $t$ and $r$) for each $l,m$-harmonic of the perturbation.
These equations couple between different tensorial components of the perturbation,
but are conveniently written in a form where no coupling occurs in the
principal part. In BS these equations are then solved in the time domain
(for each given $l,m$) using finite differentiation on a characteristic mesh.
The monopole and dipole modes ($l=0,1$) are calculated
separately, based essentially on the semi-analytical results of Detweiler and
Poisson \cite{Detweiler:2003ci}. The full perturbation is finally obtained
by summing over a sufficiently large number of modes $l,m$.

To construct the quantity $\tilde H$ in Eq.\ (\ref{final}) next requires us to obtain
the R-part of our numerical solutions, evaluated at the particle. Conveniently,
it is not strictly the R-part $\tilde h^{\rm R}_{\alpha\beta}$ that we need,
but rather the contracted quantity $\tilde h^{\rm R}_{\alpha\beta}
\tilde u^{\alpha}\tilde u^{\beta}$.
The construction of this quantity, through mode-by-mode regularization, is
prescribed by SD in Sec.\ IV.B of \cite{Detweiler:2008} for any gauge related
to the SD gauge through a $\bar\xi$-type transformation, and we employ it here
in a direct manner. (The regular gauge transformation $\hat\xi^{\alpha}=a\, t\,
\delta^{\alpha}_t$ does not affect the singular part of the perturbation,
hence the regularization procedure described in \cite{Detweiler:2008} is
applicable within the broader $\hat\xi$ family.)
We construct $\tilde H$ for a series of orbital radii $r_0$, and then use Eq.\
(\ref{final}) to obtain $\Delta U$ for these radii.
%Note that $\Delta\tilde{u}^t(\tilde R\to r_0)=\Delta\tilde{u}^t(\tilde R)
%+O(\mu^2)$ [since $\Delta\tilde{u}^t$ is itself of $O(\mu)$], so the procedure
%we have just described effectively yields numerical values for
%$\Delta\tilde{u}^t(\tilde R)$ at the desired oder.
We tabulate the values thus obtained in Table I, alongside the
corresponding SD values from Ref.\ \cite{Detweiler:2008}.
The fractional difference between the
BS and SD results is found to be similar in magnitude to the estimated fractional
numerical error in the BS data. (How this numerical error is estimated is
described in detail in Ref.\ \cite{Barack:2007tm}; the numerical error in
the corresponding SD data is in all cases much smaller.)
We conclude that the two calculations are in agreement with each other.

%We also plot the post-Newtonian
%results given in \cite{Detweiler:2008} as
%\[
% \Delta \tilde{u}^t =
% \frac{\mu}{M}\left[
% -\left( \frac{M}{R_\Omega} \right)
% -2\left( \frac{M}{R_\Omega} \right)^2
% -5\left( \frac{M}{R_\Omega} \right)^3
% -O\left( \left( \frac{M}{R_\Omega} \right)^4 \right).
% \right]
%\]

%\begin{figure}[tbp]
%\includegraphics[width=8cm]{RvsUt.eps}
%\includegraphics[width=8cm]{RvsUtup.eps}
%\caption{Correction in $\tilde{u}^t$.
%Here we plot $\Delta \tilde{u}^t$ as a function of
%$r_0$, instead of the gauge invariant radius, $R_\Omega$.
%Taking the limit of $\mu/M \to 0$, $r_0$ approaches
%to the gauge invariant radius $R_\Omega$.}
%\label{fig:RvsUt}
%\end{figure}
\begin{table}[Htb]
\begin{tabular}{c|c|c|c|c}
\hline\hline
$r_0/M$  &
$\tilde H$ &
$\Delta U$ (from BS) &
$\Delta U$ (from SD) &
Rel.\ diff. \\
\hline
$6.0$ & $-0.523602$ & $-0.296040244$ $[7e-05]$   & $-0.2960275$ & $4e-05$ \\
$6.2$ & $-0.493483$ & $-0.276743089$ $[6e-05]$   & $-0.2767327$ & $4e-05$ \\
$6.4$ & $-0.466941$ & $-0.260014908$ $[6e-05]$   & $-0.2600063$ & $3e-05$ \\
$6.6$ & $-0.443349$ & $-0.245359714$ $[6e-05]$   & $-0.2453525$ & $3e-05$  \\
$6.8$ & $-0.422222$ & $-0.232402084$ $[6e-05]$   & $-0.2323959$ & $3e-05$  \\
$7.0$ & $-0.403177$ & $-0.220852781$ $[5e-05]$   & $-0.2208475$ & $2e-05$ \\
$7.2$ & $-0.385906$ & $-0.210485427$ $[5e-05]$   & $-0.2104809$ & $2e-05$ \\
$7.4$ & $-0.370163$ & $-0.201120318$ $[5e-05]$   & $-0.2011164$ & $2e-05$ \\
$7.6$ & $-0.355745$ & $-0.192612971$ $[5e-05]$   & $-0.1926095$ & $2e-05$ \\
$7.8$ & $-0.342483$ & $-0.184845874$ $[4e-05]$   & $-0.1848428$ & $2e-05$ \\
$8.0$ & $-0.330239$ & $-0.177722443$ $[4e-05]$   & $-0.1777197$ & $2e-05$ \\
$9.0$ & $-0.280717$ & $-0.149362192$ $[3e-05]$   & $-0.1493606$ & $1e-05$ \\
$10.0$ & $-0.244630$ & $-0.129123253$ $[2e-06]$  & $-0.1291222$ & $8e-06$ \\
$11.0$ & $-0.217039$ & $-0.113875315$ $[1e-06]$  & $-0.1138747$ & $5e-06$ \\
$12.0$ & $-0.195196$ & $-0.101936046$ $[1e-06]$  & $-0.1019355$ & $5e-06$ \\
$13.0$ & $-0.177441$ & $-0.092313661$ $[1e-06]$  & $-0.09231331$& $4e-06$ \\
$14.0$ & $-0.162705$ & $-0.084382221$ $[1e-06]$  & $-0.08438195$& $3e-06$ \\
$15.0$ & $-0.150267$ & $-0.077725527$ $[1e-06]$  & $-0.07772532$& $3e-06$ \\
$16.0$ & $-0.139621$ & $-0.072055223$ $[1e-06]$  & $-0.07205505$& $2e-06$ \\
$18.0$ & $-0.122337$ & $-0.062902026$ $[1e-06]$  & $-0.06290189$& $2e-06$ \\
$20.0$ & $-0.108893$ & $-0.055827795$ $[6e-07]$  & $-0.05582771$& $2e-06$ \\
$25.0$ & $-0.085479$ & $-0.043599881$ $[3e-07]$  & $-0.04359984$& $9e-07$ \\
$30.0$ & $-0.070380$ & $-0.035778334$ $[5e-07]$  & $-0.03577831$& $7e-07$ \\
$40.0$ & $-0.052029$ & $-0.026339690$ $[3e-07]$  & $-0.02633967$& $7e-07$ \\
$50.0$ & $-0.041277$ & $-0.020844661$ $[2e-07]$  & $-0.02084465$& $5e-07$ \\
$60.0$ & $-0.034211$ & $-0.017247596$ $[1e-06]$  & $-0.01724759$& $3e-07$ \\
$70.0$ & $-0.029211$ & $-0.014709648$ $[7e-07]$  & $-0.01470964$& $5e-07$ \\
$80.0$ & $-0.025487$ & $-0.012822962$ $[6e-07]$  & $-0.01282296$& $2e-07$ \\
$90.0$ & $-0.022605$ & $-0.011365317$ $[5e-07]$  & $-0.01136531$& $6e-07$ \\
$100.0$ & $-0.020309$ & $-0.010205285$ $[4e-07]$ & $-0.01020528$& $5e-07$ \\
$120.0$ & $-0.016880$ & $-0.008475253$ $[2e-07]$ & $-0.008475251$& $2e-07$ \\
$150.0$ & $-0.013469$ & $-0.006757093$ $[3e-07]$ & $-0.006757092$& $2e-07$ \\
$200.0$ & $-0.010076$ & $-0.005050643$ $[3e-07]$ & $-0.005050642$& $2e-07$ \\
\hline\hline
\end{tabular}  \label{table}
\caption{Comparison between numerical results from BS (Barack and Sago, Ref.\
\cite{Barack:2007tm}) and SD (Detweiler, Ref.\ \cite{Detweiler:2008}).
The first column gives the radius of the circular orbit in terms of the
Schwarzschild standard areal radial coordinate $r$ [or, equivalently at the
relevant order, in terms of the gauge-invariant radius $R$---see the discussion
around Eqs.\ (\ref{almostfinal}) and (\ref{final})]. The second column
displays the numerical value of the perturbation quantity $\tilde H$ [defined in
Eq.\ (\ref{H})], as calculated in this paper using the BS Lorenz-gauge code.
In the third column we tabulate the numerical values of the gauge-invariant
$O(\mu)$ quantity $\Delta U$ [defined in Eq.\ (\ref{Deltaut})],
as calculated here based on Eq.\ (\ref{final}) and using the BS metric perturbation.
The numbers in square brackets are the estimated fractional errors in the BS data.
The fourth column displays the values of $\Delta U$ as derived
using the SD code, to be compared with the BS values in the third column.
In the fifth column we indicate the relative difference between the BS and SD
results. This difference is found to be comparable in magnitude to the estimated
numerical error in the BS data (which is larger than that in the SD data),
providing a reassuring validation test for
both analyses.}
\end{table}

%***********************************************************************
\section{Discussion}\label{Sec:V}
%***********************************************************************

The agreement established here between the numerical results of BS and SD
not only provides an important validation test for both analyses, but it
also illustrates (and confirms) a few fundamental results from GSF theory.
The following list highlights these results.
(i) The GSF, as defined and calculated by Mino, Sasaki and Tanaka \cite{Mino:1996nk}
and Quinn and Wald \cite{Quinn:1996am}, can also be derived from the
Detweiler--Whiting R-field $\tilde h_{\alpha\beta}^{\rm R}$
\cite{Detweiler:2002mi} using the formula on the RHS of Eq.\ (\ref{EOM:v4}).
(ii) The GSF can be defined and calculated through a ``same-coordinate-value''
mapping of the orbit from the physical perturbed spacetime onto a background
spacetime; relaxing the ``same-coordinate-value'' rule gives rise to the
gauge ambiguity in the GSF \cite{Barack:2001ph}.
(iii) A suitable gauge transformation can be made (here $r\to R$) which
nullifies the GSF (here, the conservative piece thereof). However, the
information about the physical finite-$\mu$ effect can then still be retrieved
in full from the metric perturbation in the new gauge---cf. Eq.\ (\ref{final}).
This is a particular example of the general statement that the full
information about the finite-$\mu$ effect is contained in the combination of
both the GSF and the metric perturbation \cite{Barack:2001ph}.

As an aside in this work, we discussed what one may mean by a ``physically
reasonable'' gauge transformation in the context of circular orbits.
Sensibly, a ``physically-reasonable'' gauge is one in which the metric perturbation
admits the approximate helical symmetry of the black hole + particle system,
i.e., it satisfies $(\partial_t + \Omega_0\partial_\phi)h_{\alpha\beta}=0$.
However, it seems unnecessary to require that the gauge transformation
generators $\xi^{\alpha}$ connecting any two such physically-reasonable gauges be
themselves helically symmetric. If fact, the general class of such
generators, denoted here $\hat\xi$, includes members $\hat\xi(a\ne 0)$ which
are {\em not} helically symmetric. The gauge transformation between the BS and SD
perturbations---both of which being ``physically-reasonable'' in the above
sense---is indeed generated by a vector $\hat\xi$ which is {\em not}
helically symmetric.

We anticipate that comparisons similar to the one discussed in this work will
allow robust tests of self-force calculations for other orbits and other
spacetimes (e.g., Kerr) when such calculations are available. The essential
elements of our formal discussion are directly applicable to other orbits and
geometries. Most important, Eqs.\ (\ref{tauratio}) and (\ref{uBS}), which
describe the mapping of the orbital elements from a BS-type background spacetime
to an SD-type perturbed spacetime, hold quite generally for any orbit in any
black hole spacetime, and could form a basis for future comparisons.
The major challenge in any such future comparison would remain to devise
a suitable set of gauge-invariant quantities.

\section*{ACKNOWLEDGEMENTS}

LB and NS acknowledge support from PPARC/STFC through grant number PP/D001110/1.
SD acknowledges support from National Science Foundation, through
grant number PHY-0555484. NS thanks the Yukawa Institute for Theoretical
Physics for hospitality during the initial stage of his work on this project.
We thank the participants of the 11th Capra meeting in Orl\'{e}ans for useful
discussions.

\end{document}